# THE DEMYSTIFICATION OF EMERGENT BEHAVIOR

*Gerald E. Marsh*

Argonne National Laboratory (Ret)

5433 East View Park

Chicago, IL 60615

E-mail: gemarsh@uchicago.edu

**Abstract.** Emergent behavior that appears at a given level of organization may be characterized as arising from an organizationally lower level in such a way that it transcends a mere increase in the behavioral degree of complexity. It is therefore to be distinguished from systems exhibiting chaotic behavior, for example, which are deterministic but unpredictable because of an exponential dependence on initial conditions. In emergent phenomena, higher-levels of organization are not determined by lower-levels of organization; or, more colloquially, emergent behavior is often said to be "greater than the sum of the parts". The concept plays an especially important but contentious role in the biological sciences. This essay is intended to demystify at least some aspects of the mystery of emergence.

Complex systems, and in particular biological systems, often display what has come to be known as emergent behavior. Associated with this phenomenon is a sense of the mysterious: the emergent properties of the collective whole do not in any transparent way derive from the underlying rules governing the interaction of the system's components. Unfortunately, there is not even a universally acknowledged definition of emergence. Nor do the concept and its explication in the literature constitute an organized, rigorous theory. Instead, it is more of a collection of ideas that have in common the notion that complex behavior can arise from the underlying simple rules of interaction. There are, however, some very interesting attempts at definition and theory. See, for example, Ryan[1] and references cited therein, and the book edited by Clayton and Davies.[2]

The philosophical literature[3] separates emergence into two basic categories: *strong emergence*, where higher-level processes cannot *in principle* be derived from lower-level processes; and *weak emergence*, where there is no known way to derive higher-level processes from lower-level processes, but it has not been shown that this cannot be done. It should be noted, as should become clear in what follows, that an acceptance of the concept of emergence does not imply that a denial of reductionism must follow. It does however mean we must be careful about what we mean by reductionism.

Ernst Mayr[4] in his monumental work *The Growth of Biological Thought* characterizes emergence as follows:

> "Systems almost always have the peculiarity that the characteristics of the whole cannot (not even in theory) be deduced from the most complete knowledge of the components, taken separately or in other partial combinations. This appearance of new characteristics in wholes has been designated as *emergence*. Emergence has often been invoked in attempts to explain such difficult phenomena as life, mind, and consciousness. Actually, emergence is equally characteristic of inorganic systems. As far back as 1868, T. H. Huxley asserted that the peculiar properties of water, its 'aquosity,' could not be deduced from our understanding of the properties of hydrogen and oxygen. The person, however, who was more responsible than anyone else for the recognition of the importance of emergence was Lloyd Morgan. There is no question, he said, 'that at various grades of organization, material configurations display new and unexpected phenomena and that these include the most striking features of adaptive machinery.' Such emergence is quite universal and, as Popper said, 'We live in a universe of' emergent novelty'. Emergence is a descriptive notion which, particularly in more complex systems, seems to resist analysis. Simply to say, as has been [done][†], that emergence is due to complexity is, of course, not an explanation. Perhaps the two most interesting characteristics of new wholes are (1) that they, in turn, can become parts of still higher-level systems, and (2) that wholes can affect properties of components at lower levels. The latter phenomenon is sometimes referred to as 'downward causation'. Emergentism is a thoroughly materialistic philosophy. Those who deny it, like Rensch, are forced to adopt pan-psychic or hylozoic theories of matter." [references deleted]

While the concept of emergent behavior may be difficult to define, it is easy to intuitively understand. One defines a set of rules for the interaction of a class of objects with each other and with the environment, and within the constraints set by the rules, the resulting observed behavior is often very complex, transcending the simplicity of the rules themselves. A concrete and practical example is the class of artificial insects developed at MIT,[5, 6] particularly the emergent behavior displayed when central control modules are eliminated and the robots are allowed to "learn" their behavior.

Randall Beer, Hillel Chiel, and Leon Sterling have done similar work on simulating insect-like neural networks on a digital computer at Case Western Reserve University.[7] The body plan they used is based on the American cockroach and the artificial insect they created walks with changes in gait that are an emergent property similar to the changes in gait made by real insects as they change their speed. That is, these gait changes are not "wired into" the circuitry. This complex behavior is achieved with only six "neurons" per leg, and two command neurons that are wired to all six legs. A six-legged robot duplicating the neural circuitry used in the computer simulation was constructed and it displayed a range of gaits similar to those of the simulated insect.

Deborah Gordon has given another example of emergence[8] in her studies of the complex behavioral patterns exhibited by ant colonies. These behaviors, and the changes in behavior as the colony grows and ages, seems to be based on the decisions of individual ants that operate with a relatively simple set of rules based on social contact and the environment within which the individual finds itself.

---

[†] Original has the word "clone".

The various tasks carried out in an ant colony are accomplished without direction, without the chain of command implicit in an hierarchical organization. A single set of rules, at the level of the individual, is responsible for the behavior of the colony as a whole. As an ant colony ages, individual responses are refined to adapt to changes in the colony's environment. Such refinement represents a type of social learning, with the memory trace being stored in a holographic fashion across the individuals of the colony. Small modifications of the rules of interaction can lead to significant changes in the behavior of the colony as a whole. Of course this discussion is oversimplified in that real ant colonies are far more complex and the inhabitants of ant colonies have specialties so that there may be several subsets of rules rather than a single overarching set. There exist many other examples of emergence.

The emergent behavior of a system, while it is determined by the elements of the system and the rules of interaction between them—and perhaps with the environment, is not contained explicitly in any of the rules or elements themselves, nor is the behavior explained by a simple summation over the components making up the system. Emergent behavior is characterized by being "greater than the sum of the parts."

While emergent behavior is completely dependent on the set of rules governing the interaction between the elements of a system, a key question is whether, at the level of the emergent behavior, new rules of interaction appear that are not, in a fundamental sense, predictable. Very simple mathematical models can exhibit extremely complex dynamics even though the behavior is completely deterministic. They have been successfully used to model the dynamics of systems in a variety of fields, including electronics, mechanics, and biology.[9] In each of these diverse areas, even when the behavior is chaotic (where the dynamics depends exponentially on changes in initial conditions), and the dynamical trajectories may look like random noise, the behavior is deterministic. In terms of emergence, one might characterize such chaotic, deterministic behavior as perhaps fitting into the category of weak emergence.

Mathematics can also provide examples of truly emergent properties. A Möbius strip, for example, is a one-sided non-orientable surface with one edge. The one-sidedness of a Möbius strip only exists if it is not cut. If it is, the strip becomes two sided. One might say that the one-sidedness is an emergent global property of the complete Möbius strip.

The higher up one goes in a given hierarchy of emergent behavior, the more the organization seems completely independent of the rules determining the behavior of the levels below—which, nevertheless, is not to deny that the higher-order rules are in some sense inherently determined by the properties of the component parts.[†] But it is the definition of "inherently determined" that contains the essence of the problem.

How can one resolve this conundrum? The answer may lie in the new, internal degrees of freedom that appear as one ascends a hierarchy of emergence. Consider first a simple example from elementary classical mechanics that has relevance to the formation of molecules and hence also to biology. The number of positional degrees of freedom for 2N particles is given by the product of the number of particles and the number of coordinates needed to specify the location of each of the particles in 3-dimensional space. This is $2N \times 3 = 6N$. Now if the particles are combined so as to produce N bonded pairs, with some bonding distance associated with each pair, the number of external degrees of freedom for the pairs is reduced to 3N. However, new degrees of freedom internal to each of the pairs have appeared—the distance between the particles constituting each pair, and the two angles needed to specify the orientation of each pair in 3-dimensional space, a total of three internal degrees of freedom. Notice that the total number of degrees of freedom (3N to locate the pairs in 3-space and 3N "emergent", internal degrees of freedom) has remained constant.

If we had started with 3N particles having 9N degrees of freedom and combined them to form N linear chain triplets, the original number of degrees of freedom would be reduced to 3N but the emergent internal degrees of freedom would add up to 6N so that the sum remains 9N. For a linear chain configuration, the internal degrees of freedom are the two distances between the particles constituting each triplet, the two angles needed to orient the line between the first and second particle, and the two angles needed to orient the line from the second to third particles, a total of six for each triplet. For all the triplets, the total number of internal degrees of freedom would then be 6N. Adding the 3N degrees of freedom needed to locate the N triplets in 3-space gives a grand total of 9N. If the configuration of the triplets is changed to a triangular configuration, the definition of the internal degrees of freedom will also change: for each triplet the internal degrees of freedom would be the three distances between the particles comprising each triplet and the three angles needed to orient the triangular configuration in 3-space (two angles to orient the normal to the configuration and one angle for rotation about the normal).

---

[†] Mayr's discussion of "Explanatory Reductionism" is relevant here.

The concept of degrees of freedom comes from classical mechanics where the number of degrees of freedom may be reduced because of the existence of constraints. If a system with $n$ degrees of freedom having coordinates $q_1$, $q_2$, …$q_n$ is subject to $k$ constraint equations having the form $f_r(q_1, q_2, … q_n) = 0$, $r = 1, 2, … k$, these equations can be solved for $k$ of the coordinates in term of the remaining $n–k$. The resulting $n–k$ coordinates may be varied independently without violating the constraints. The system thus has $n–k$ degrees of freedom and $q_1$, $q_2$, …$q_{n-k}$ may be taken as generalized coordinates, the constraints having been eliminated. An example might be helpful.

Consider a single particle free to move in 3-dimensional space and thus having three degrees of freedom. Now constrain the particle to move on a surface given by $f(x,y,z) = 0$. This is the constraint equation, and it can be solved for $z$ in terms of $x$ and $y$, which become the generalized coordinates. The constrained particle now only has two degrees of freedom. The extra degree of freedom, unlike the composite particle examples given above, is eliminated by the constraint equation. Constraints could, of course, also exist for a composite particle in which case some of the internal degrees of freedom could be eliminated using the same procedure given above.

The conservation of the number of degrees of freedom is subtler in quantum mechanics. Take, for example, the case of the helium atom. A helium atom is comprised of a nucleus (considered as a single particle) and two electrons. As separate particles, assumed to be localized in space, the number of degrees of freedom for the three particles is nine. The combination of the three particles to form a helium atom would lead to three degrees of freedom for the location of the nucleus and six additional degrees of freedom consisting of the quantum numbers $n$, $l$, and $m$ for each of the electrons. Of course, this is not the whole story since quantum mechanics sets additional constraints on the numerical values of the quantum numbers $n$, $l$ and $m$.

What one may call emergent behavior already appears even at the lowest level of reductionist exploration achieved to date. Ordinary matter derives its mass from the kinetic and potential energy of the massless gluons and nearly massless quarks that make up protons, neutrons, and therefore all atomic nuclei. The theory (which can be viewed as a set of rules) governing their behavior is called quantum chromodynamics or QCD. Unlike many other theories, QCD has no adjustable parameters other than its overall coupling strength. Nonetheless, it has not proved possible to describe nuclei in terms of their fundamental quark and gluon constituents. Instead, so-called "effective" models of nuclei have been developed: in high-energy physics, nuclei are treated as a collection of free quarks; in the low-energy regime, on the other hand, one uses the many-body and shell models.

As one lowers the energy scale to the point where the interaction between distinct protons and neutrons becomes important, new internal degrees of freedom appear that are not contained in the higher energy theory. One might well view the properties of the low-energy regime as being emergent. Although it is still not clear whether one should consider this a case of weak or strong emergence.

Similarly, the structure and variety of all atoms are determined by the rules of quantum mechanics. But the form of the lattice they or their compounds form may depend on emergent degrees of freedom such as temperature and pressure reflecting environmental factors. Although one might argue the varieties of structural forms depending on such environmental factors are emergent, it is also possible to argue—at least in principle—that the underlying quantum mechanical rules could take them into account.

An example where this is not possible is the chemistry of saturated hydrocarbons. The rules of quantum mechanics certainly determine the bonding of carbon and hydrogen, and no matter how structurally complex the hydrocarbon, these rules are faithfully obeyed. But the rules of quantum mechanics say nothing about how many carbon atoms may form a chain or whether they form straight chains or branched-chain carbon skeletons. There are emergent degrees of freedom that appear when atoms combine to form these hydrocarbon molecules. It is these emergent degrees of freedom that determine the chemical properties of the saturated hydrocarbons and these chemical properties could well be viewed as an emergent property of a complex system (saturated hydrocarbon molecules) not fully determined by the underlying quantum mechanical rules governing the bonding of hydrogen and carbon atoms.

The emergent rules that govern the chemistry of saturated hydrocarbons are *dependent* on the underlying rules governing the bonding of hydrogen and carbon, but are not *determined* by these rules. That is, the emergent rules cannot be derived from the underlying quantum mechanical rules governing hydrogen-carbon bonding. The difference here is similar to that found in mathematics between *necessity* and *sufficiency*. It is this distinction that should be used to inform the definition of reductionism, particularly in biology with its hierarchical organization, in light of the reality of emergent phenomena.

In the same way, the rules of chemical bonding (again reflecting the rules of quantum mechanics) specify the structure of DNA, but not the sequence of bases. The possible sequences of bases and length of the DNA molecule itself again constitute emergent degrees of freedom not specified by the rules of chemical bonding. The sequence of bases determines the genes coding for proteins, small RNAs, etc., and it is roughly at this level that the environment begins to play a significant role in the evolution of life through the Darwinian process of variation and selection.[†] But it is not only the set of genes that is responsible for the diversity of animal forms. Of primary importance are differences in gene regulation during ontogeny.[10]

The sequence of bases in DNA contains regulatory code that governs gene expression both in time and location. While this code constitutes another higher-level set of rules, they are rules that have the additional property of being able to change with time in response to environmental selection at the organismal level. The emergent number of degrees of freedom appearing at this level vastly exceeds those at lower hierarchic levels. The whole issue of epigenetics—defined as heritable changes in gene expression not due to changes in base sequence, essentially what allows cells having the same genetic inheritance to make up the variety of cell types comprising an organism—and its role in evolution is still an active area of research.[11, 12]

---

[†] In terms of the origin of life, molecular evolution and the earliest living creatures were of course also subject to Darwinian variation and selection.

At an even higher level, while DNA surely determines the structure of living creatures, it would be impossible to derive their social behavior and organization from only the sequence of bases in DNA.

The hierarchy above, starting from the lowest reductionist level now known has led to the rules governing ontogeny. Indeed, from an Olympian point of view, life itself may be viewed as an emergent property of matter. But the hierarchy does not apparently end and may well be truly open in the sense of Köestler.[13]

Consciousness and intelligence appear to emerge gradually as the complexity of life increases. Simultaneously, and as a parallel development, a social structure comes into existence. Social behavior can be as simple as that of slime molds when forming a fruiting body, be relatively complex as in the behavior of an ant colony, or be represented by the far more complex behavior of human societies. All appear as forms of emergent behavior.

If the idea that emergent behavior results from the coming into being of new, internal degrees of freedom that arise as one ascends a given hierarchy of emergence is to hold, the inverse should also be true in the sense that a reductionist analysis should eliminate degrees of freedom in the process of descending the hierarchy through reductionist analysis. Here, reductionism is defined as gaining an understanding of a complex system through detailed analysis of the components of the system and their interactions. From the examples of emergence given above, this would seem to be almost trivially true.

In sum, one should view emergence and reductionism as opposite sides of the same coin. Dissecting complex behavior from the top down eliminates internal degrees of freedom in the course of analysis, while emergent phenomena occur when internal degrees of freedom appear when combining component elements into more complex systems. If individual ants are studied to determine their rules of interaction, there is nothing mysterious about the process. But given those rules, one cannot predict the behavior of the colony because the new degrees of freedom that appear in the collective colony cannot be deduced from the rules of interaction—these rules are *necessary* but not *sufficient* to predict the emergent behavior. It is the unexpected consequences of the additional degrees of freedom that appear mysterious.